\documentclass[amsmath,amssymb,aps,10pt,prd,twocolumn,letterpaper,nofootinbib,balancelastpage,notitlepage,superscriptaddress,floatfix,preprintnumbers]{revtex4-2}
\usepackage{graphicx}	% Standard figures 
\usepackage{ragged2e}	% for justified text in JHEP.bst
\usepackage{bm}		    % Bold math
\usepackage{tikz}		% Tikz figures
\usetikzlibrary{math}
\usepackage[sort&compress]{natbib}	 	% Standard reference package
	\setcitestyle{square,numbers,comma}	% Set citation style
\usepackage[colorlinks=true,urlcolor=blue,linkcolor=red,citecolor=red]{hyperref}	
\newcommand{\figref}[2][]{Fig{#1}.~\ref{fig:#2}}		% Figure reference
\newcommand{\secref}[2][]{Sec{#1}.~\ref{sec:#2}}		% Section reference
\newcommand{\appref}[2][x]{Appendi{#1}~\ref{app:#2}}	% Appendix reference
\renewcommand{\eqref}[2][]{Eq{#1}.~(\ref{eq:#2})}		% Equation reference
\newcommand{\citeR}[2][]{Ref{#1}.~\cite{#2}}			% Ref. Citation
\newcommand{\rh}{\mathrm{rh}}
\begin{document}
%%%%%%%%%%%%%%%%%%%%%%%%%%%%%%%%%%%%%%%%%%%%%%%%%%%%%%%%%%%%%%%%%%%%%%%%%%%%%%%%%%%%%%%%%%
%%%%%%%%%%%%%%%%%%%%%%%%%%%%%%%%%%%%%%%%%%%%%%%%%%%%%%%%%%%%%%%%%%%%%%%%%%%%%%%%%%%%%%%%%%
%%%%%%%%%%%%%%%%%%%%%%%%%%%%%%%%%%%%%%%%%%%%%%%%%%%%%%%%%%%%%%%%%%%%%%%%%%%%%%%%%%%%%%%%%%
%%%%%%%%%%%%%%%%%%%%%%%%%%%%%%%%%%%%%%%%%%%%%%%%%%%%%%%%%%%%%%%%%%%%%%%%%%%%%%%%%%%%%%%%%%

%%%%%%%%%%%%%%%%%%%%%%%%%%%%%%%%%%%%%%%%%%%%%%%%%%%%%%%%%%%%%%%%%%%%%%%%%%%%%%%%%%%%%%%%%%
% Title, Author and Affiliation
\title{Dynamically generated tilt of isocurvature fluctuations}
\date{\today}
%%%%%%%%%%%%%%%%%%%%%%%%%%%%%%
\author{Saarik Kalia}
\email{skalia@ifae.es}
\affiliation{School of Physics \& Astronomy, University of Minnesota, Minneapolis, MN 55455, USA}
\affiliation{Institut de F\'isica d’Altes Energies (IFAE), The Barcelona Institute of Science and Technology, Campus UAB, 08193 Bellaterra (Barcelona), Spain}
%%%%%%%%%%%%%%%%%%%%%%%%%%%%%%%%%%%%%%%%%%%%%%%%%%%%%%%%%%%%%%%%%%%%%%%%%%%%%%%%%%%%%%%%%%

%%%%%%%%%%%%%%%%%%%%%%%%%%%%%%%%%%%%%%%%%%%%%%%%%%%%%%%%%%%%%%%%%%%%%%%%%%%%%%%%%%%%%%%%%%
\preprint{UMN-TH-4510/25}
%%%%%%%%%%%%%%%%%%%%%%%%%%%%%%%%%%%%%%%%%%%%%%%%%%%%%%%%%%%%%%%%%%%%%%%%%%%%%%%%%%%%%%%%%%

%%%%%%%%%%%%%%%%%%%%%%%%%%%%%%%%%%%%%%%%%%%%%%%%%%%%%%%%%%%%%%%%%%%%%%%%%%%%%%%%%%%%%%%%%%
% Abstract
\begin{abstract}
%%%%%%%%%%%%%%%%%%%%%%%%%%%%%
Light scalar fields acquire isocurvature fluctuations during inflation.  While these fluctuations could lead to interesting observable signatures at small scales, they are strongly constrained on large scales by cosmic microwave background observations.  When the mass of the scalar is much lighter than the inflationary Hubble scale, $m\ll H_I$, the spectrum of these fluctuations is flat.  Meanwhile, if $m\gg H_I$, the fluctuations are suppressed.  A blue-tilted isocurvature spectrum which exhibits enhanced structure on small scales but avoids observational constraints on large scales therefore requires a coincidence of scales $m\sim H_I$ for a free massive scalar.  In this work, we show that if a scalar field possesses a nontrivial potential, its inflationary dynamics naturally cause this condition to be satisfied, and so a blue-tilted spectrum is generically expected for a large class of potentials.  Specifically, if its potential $V$ exhibits a region which satisfies the slow-roll condition $V''<3H_I^2$, the scalar condensate will spend most of inflation close to the boundary of this region, so that its effective mass is typically close to $H_I$.  The resulting blue tilt is inversely proportional to the number of $e$-folds of inflation prior to horizon crossing.  If the scalar is long-lived, this mechanism leads to an attractor prediction for its relic abundance, which is insensitive to initial conditions of the scalar.  In particular, a scalar field with quartic self-interactions can achieve the correct abundance to constitute all of the dark matter for a wide range of masses.  We compute the relationship between the mass and self-coupling of quartic dark matter predicted by this mechanism.
%%%%%%%%%%%%%%%%%%%%%%%%%%%%%
\end{abstract}
%%%%%%%%%%%%%%%%%%%%%%%%%%%%%%%%%%%%%%%%%%%%%%%%%%%%%%%%%%%%%%%%%%%%%%%%%%%%%%%%%%%%%%%%%%

%%%%%%%%%%%%%%%%%%%%%%%%%%%%%%%%%%%%%%%%%%%%%%%%%%%%%%%%%%%%%%%%%%%%%%%%%%%%%%%%%%%%%%%%%%
\maketitle
%%%%%%%%%%%%%%%%%%%%%%%%%%%%%%%%%%%%%%%%%%%%%%%%%%%%%%%%%%%%%%%%%%%%%%%%%%%%%%%%%%%%%%%%%%

%%%%%%%%%%%%%%%%%%%%%%%%%%%%%%%%%%%%%%%%%%%%%%%%%%%%%%%%%%%%%%%%%%%%%%%%%%%%%%%%%%%%%%%%%%
%%%%%%%%%%%%%%%%%%%%%%%%%%%%%%%%%%%%%%%%%%%%%%%%%%%%%%%%%%%%%%%%%%%%%%%%%%%%%%%%%%%%%%%%%%
\section{Introduction}
\label{sec:introduction}
%%%%%%%%%%%%%%%%%%%%%%%%%%%%%%%%%%%%%%%%%%%%%%%%%%%%%%%%%%%%%%%%%%%%%%%%%%%%%%%%%%%%%%%%%%
%%%%%%%%%%%%%%%%%%%%%%%%%%%%%%%%%%%%%%%%%%%%%%%%%%%%%%%%%%%%%%%%%%%%%%%%%%%%%%%%%%%%%%%%%%

Although the Standard Model (SM) of particle physics explains many observed phenomena, a number of open problems necessitate the existence of new physics beyond the SM.  Perhaps the simplest possible modification to the SM is the addition of a new scalar field.  Most notably, if such a new field is cosmologically stable, it could constitute the dark matter (DM)~\cite{Preskill:1982cy,Abbott:1982af,Dine:1982ah,Peebles_1999,Arias_2012,Arvanitaki_2015,Brzeminski_2021,Cyncynates_2025}.  Scalar fields also appear in several theoretical models designed to address other open problems, including the strong CP problem~\cite{Peccei:1977hh,Weinberg:1977ma,Wilczek:1977pj}, the hierarchy problem~\cite{Gherghetta_1996,Martin:1997ns,Graham_2015}, the origin of the matter-antimatter asymmetry~\cite{AFFLECK1985361,Cohen:1988kt,Co_2021}, and the cosmological constant problem~\cite{Abbott:1984qf,Ford_1987,Alberte_2016,Graham_2019}.

Extensive cosmological observations indicate that the universe is nearly homogeneous, with correlated (adiabatic) primordial fluctuations $P^\mathrm{ad}_\delta\sim10^{-9}$ in all components~\cite{planck}.  In order to explain this observation, a period of early expansion, known as inflation, is often proposed~\cite{Guth_1981,Linde:1981mu}.  One generic feature of scalar fields is that they can acquire independent (isocurvature) fluctuations during inflation.  Specifically, a free scalar field with mass much smaller than the inflationary Hubble scale, $m\ll H_I$, acquires isocurvature fluctuations on all scales, while the fluctuations of a heavy scalar with $m\gg H_I$ are heavily suppressed~\cite{Birrell:1982ix,Lyth_2002}.  On small scales, isocurvature fluctuations can lead to interesting signatures, such as gravitational waves~\cite{Domenech_2022,Ebadi:2023xhq,GarciaVerner_2025} or non-Gaussianities~\cite{Linde_1997,Kawasaki:2008pa,Kalia_2025}.  However, they are constrained to have $P^\mathrm{iso}_\delta\lesssim10^{-10}$ on large scales $k\lesssim10^{-2}\,\mathrm{Mpc}^{-1}$~\cite{planck_inflation}.  In order for a new non-interacting scalar field to present interesting signatures from enhanced fluctuations, it must therefore have $m\sim H_I$, which results in a blue-tilted isocurvature spectrum~\cite{Chung_2015,riotto2017,baumann2018,Ebadi:2023xhq}.  Lighter scalars may gain a large effective mass $m_\mathrm{eff}$ during inflation via non-minimal couplings to gravity or interactions with other fields, in order to avoid isocurvature constraints~\cite{Jeong_2013,Garcia_2023,GarciaKe_2025,Graham:2025iwx}.  By contrast, a non-interacting vector field naturally exhibits a blue-tilted spectrum~\cite{Graham_2016}.

\begin{figure*}[t]
\includegraphics[width=0.49\textwidth]{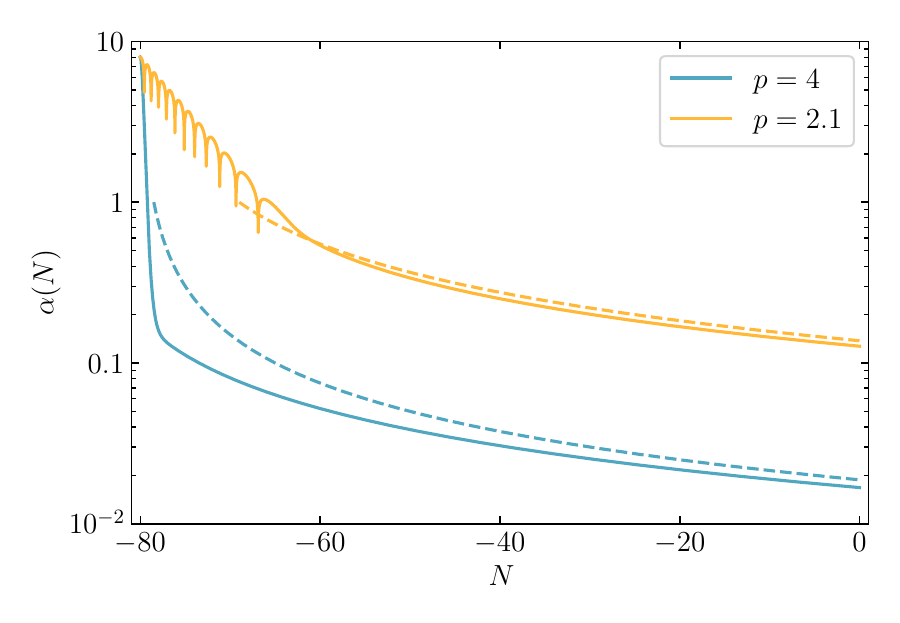}
\includegraphics[width=0.49\textwidth]{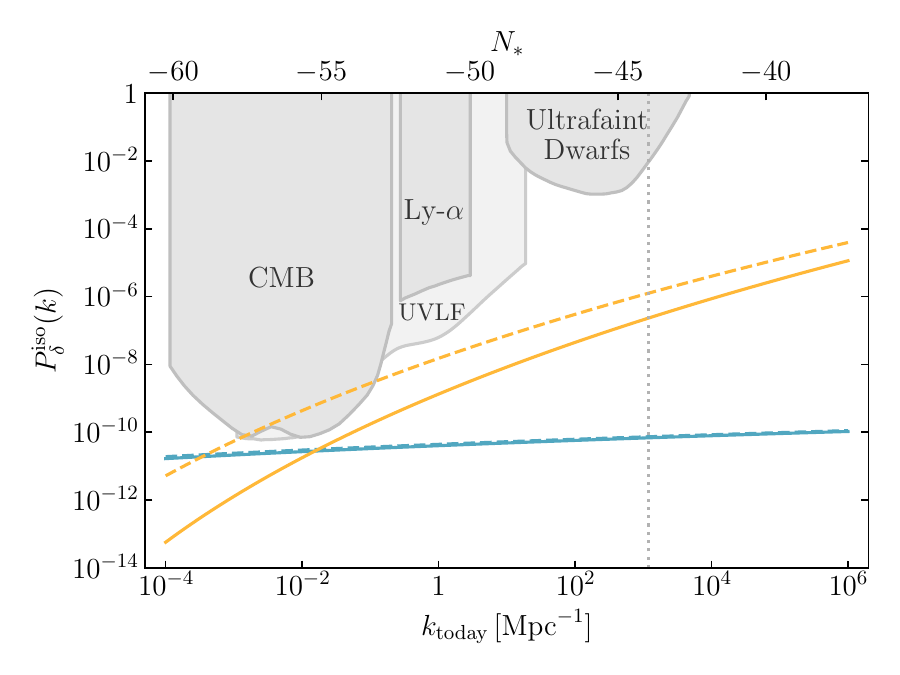}
\caption{\label{fig:Pdel}%
    \emph{Left:} Evolution of $\alpha$ [see \eqref{alpha}] as a function of $e$-folds $N$ during inflation.  We take the inflationary Hubble scale to be $H_I=10^{12}\,\mathrm{GeV}$ in this plot.  The spectator begins with $\alpha_i\gg1$ at the start of inflation $N_i=-80$.  Initially, $\alpha(N)$ oscillates and decreases in magnitude exponentially.  Once $\alpha<1$, it decreases as $1/N$.  Dashed lines show the analytic approximation in \eqref{alpha_est}.  We show evolutions for two choices of potential $V(\phi)\sim\phi^p$.  The blue curve has $\kappa=\mathcal O(1)$, while the orange has small $\kappa$.  In accordance with \eqref{alpha_est}, the latter results in much larger values of $\alpha$. \emph{Right:} Primordial isocurvature spectrum for scalar DM with same potentials as in left plot.  Dashed lines show the analytic approximation in \eqref{Pdel}.  In both numerical and analytic curves, we include the constant factor from late-time evolution [see \eqref{primordial}], which for quartic DM with mass $m=10\,\mathrm{eV}$ applies to the left of the grey dotted line [see \eqref{sep_bound}].  The tilt of the spectrum is related to $\alpha$ at horizon crossing $N_*$ (shown on top axis), as in \eqref{tilt}.  For $\kappa=\mathcal O(1)$, this generically results in a small blue tilt.  For smaller $\kappa$, as in the orange curve, the tilt can be much stronger.  In grey, we show constraints on the primordial isocurvature spectrum from observations of the CMB~\cite{Buckley_2025}, ultraviolet luminosity function (ULVF)~\cite{Co_2026}, Lyman-$\alpha$ forest~\cite{Bird_2011,Graham_2024}, and ultrafaint dwarf galaxies~\cite{Graham_2024} (see also \citeR{Gorghetto_2025}).  In order to avoid these constraints, the normalizations of the potentials are fixed at $\lambda=10^{-9}$ [as in \eqref{quartic}] for the blue curve and $\Lambda=4.4\times10^{11}\,\mathrm{GeV}$ [as in \eqref{potential}] for the orange curve.  Note that the total matter power spectrum will be the sum of the isocurvature and adiabatic fluctuations [see \eqref{primordial}].}
\end{figure*}

The purpose of this work is two-fold.  The first purpose is to demonstrate that self-interactions can generically result in a blue-tilted spectrum for a scalar spectator $\phi$ during inflation.  Specifically, if $\phi$ possesses a nontrivial potential $V(\phi)$ and inflation does not last too long, the resulting spectrum will exhibit a blue tilt, for a large class of potentials.  Similar scenarios have been studied in the case where inflation lasts a long time, so that the condensate (or zero mode) $\phi_0$ settles near the minimum of its potential~\cite{Starobinsky:1986fx,Markkanen_2019,GarciaKe_2025}.  In our scenario, the condensate will instead spend the majority of inflation undergoing slow-roll dynamics~\cite{Dimopoulos_2003,Harigaya_2013}.  In particular, if $\phi_0$ begins outside the regime where the slow-roll approximation holds, but rolls toward it, then it will quickly enter the slow-roll regime and remain near its boundary for the duration of inflation.  In other words, the slow-roll condition will be nearly saturated for the majority of inflation.  As we will see, the slow-roll condition is precisely $m_\mathrm{eff}^2=V''(\phi_0)\lesssim H_I^2$.  Therefore, the dynamics of the condensate naturally produce the conditions necessary for a blue-tilted spectrum!  We will see that the tilt of the spectrum is inversely proportional to the number of $e$-folds before horizon crossing, i.e. if inflation began $\mathcal O(10)$ $e$-folds before the cosmic microwave background (CMB) modes exited the horizon, a generic potential will result in an $\mathcal O(0.1)$ blue tilt at CMB scales.  Particular choices of potential can result in larger tilts (see \figref{Pdel}).

The second purpose is to demonstrate that if $\phi$ is long-lived, this mechanism can produce the correct relic abundance for $\phi$ to constitute the DM.  Moreover, because the slow-roll solution is an attractor, this relic abundance is insensitive to initial conditions of the scalar field.  For fixed inflationary parameters, this implies a predictive relationship between the parameters of the potential.  Unlike mechanisms such as misalignment production of axion DM, this relationship is robust to the stochastic initial value of $\phi$~\cite{Arias_2012,Graham_2018}.  In \figref{quartic}, we show this relationship between the mass $m$ and self-coupling $\lambda$ of a quartic potential.

This work is organized as follows.  In \secref{inflation}, we study the dynamics of $\phi$ during inflation, focusing first on its condensate and then on its perturbations.  We derive the slow-roll condition for the condensate, and show that the slow-roll parameter controls the tilt of the perturbation spectrum.  In \secref{radiation}, we study the subsequent dynamics during radiation domination, in the case where $\phi$ makes up the DM.  In this section, we choose to focus on a quartic potential [see \eqref{quartic}].  We first track the relic abundance, and then see how the perturbations derived in \secref{inflation} evolve after inflation.  In \secref{conclusion}, we conclude.  In \appref{relic}, we elaborate on the details of the relic density calculation, and in \appref{perturbations}, we study how the density perturbations of a scalar field evolve.  We make all the code used in this work publicly available on Github~\cite{github}.

%%%%%%%%%%%%%%%%%%%%%%%%%%%%%%%%%%%%%%%%%%%%%%%%%%%%%%%%%%%%%%%%%%%%%%%%%%%%%%%%%%%%%%%%%%
%%%%%%%%%%%%%%%%%%%%%%%%%%%%%%%%%%%%%%%%%%%%%%%%%%%%%%%%%%%%%%%%%%%%%%%%%%%%%%%%%%%%%%%%%%
\section{Inflation}
\label{sec:inflation}
%%%%%%%%%%%%%%%%%%%%%%%%%%%%%%%%%%%%%%%%%%%%%%%%%%%%%%%%%%%%%%%%%%%%%%%%%%%%%%%%%%%%%%%%%%
%%%%%%%%%%%%%%%%%%%%%%%%%%%%%%%%%%%%%%%%%%%%%%%%%%%%%%%%%%%%%%%%%%%%%%%%%%%%%%%%%%%%%%%%%%

%%%%%%%%%%%%%%%%%%%%%%%%%%%%%%%%%%%%%%%%%%
%%%%%%%%%%%%%%%%%%%%%%%%%%%%%%%%%%%%%%%%%%
%%%%%%%%%%%%%%%%%%%%%%%%%%%%%%%%%%%%%%%%%%
%%%%%%%%%%%%%%%%%%%%%%%%%%%%%%%%%%%%%%%%%%

We begin by studying the evolution of a spectator scalar field $\phi$ during inflation.  We will consider the scenario where the spectator consists of small perturbations on top of a homogeneous condensate%
%%%%%%%%%%%%
\footnote{Note that even if the scalar field begins in an inhomogeneous configuration, inflation will effectively ``zoom in" on a small region of the configuration, leading to a homogeneous scenario.  More specifically, the fluctuations of the field over a Hubble patch are $\delta\phi\sim(aH_I)^{-1}\nabla\phi$.  Even in the absence of dynamics of the field, this redshifts relative to $\phi_0$, so that \eqref{homogeneous} will eventually become a valid perturbative expansion.}
%%%%%%%%%%%%
\begin{equation}\label{eq:homogeneous}
    \phi(x,t)=\phi_0(t)+\delta\phi(x,t).
\end{equation}
First, we will study the dynamics of the condensate during inflation in the presence of a nontrivial potential~$V(\phi)$.  We will derive the slow-roll condition%
%%%%%%%%%%%%
\footnote{The parameter $\alpha$ is analogous to the second slow-roll parameter $\eta$ of the inflaton (by substituting $H^2=V_{\inf}/3M_\mathrm{pl}^2$).  Its behavior is however very different!  For a monomial inflaton potential, $\eta$ increases as inflation progresses.  Meanwhile, for a monomial spectator potential, $\alpha$ decreases over time [see \eqref{alpha_sol}].}
%%%%%%%%%%%%
\begin{equation}\label{eq:alpha}
    |\alpha|\equiv\frac{|V''(\phi_0)|}{3H^2}\ll1
\end{equation}
for the condensate dynamics (where $H=\dot a/a$ is the Hubble scale), and see that for a large class of potentials, $\alpha\lesssim1$ for most of inflation.  Then we will study the perturbations of the spectator.  In particular, we will show that the spectrum of the perturbations receive a blue tilt from a nonzero $\alpha$.  We therefore conclude that a small blue tilt can generically arise in the power spectrum of a spectator.

Throughout this section, we will ignore the dynamics of the inflaton, and we will assume an exact de Sitter spacetime for inflation, given by the scale factor
\begin{equation}
    a=e^{H_It}=e^N=-\frac1{H_I\tau},
\end{equation}
in terms of physical time $t$, $e$-folds $N$, or conformal time~$\tau$.  This corresponds to a constant Hubble scale $H(t)=H_I$.  Accounting for the evolution of $H(t)$ introduces subleading corrections to the tilt of the power spectrum (see footnote \ref{ftnt:slowroll}).

%%%%%%%%%%%%%%%%%%%%%%%%%%%%%%%%%%%%%%%%%%%%%%%%%%%%%%%%%%%%%%%%%%%%%%%%%%%%%%%%%%%%%%%%%%
%%%%%%%%%%%%%%%%%%%%%%%%%%%%%%%%%%%%%%%%%%%%%%%%%%%%%%%%%%%%%%%%%%%%%%%%%%%%%%%%%%%%%%%%%%
\subsection{Condensate}
\label{sec:condensate}
%%%%%%%%%%%%%%%%%%%%%%%%%%%%%%%%%%%%%%%%%%%%%%%%%%%%%%%%%%%%%%%%%%%%%%%%%%%%%%%%%%%%%%%%%%
%%%%%%%%%%%%%%%%%%%%%%%%%%%%%%%%%%%%%%%%%%%%%%%%%%%%%%%%%%%%%%%%%%%%%%%%%%%%%%%%%%%%%%%%%%

The evolution of the spectator condensate during inflation is given by%
%%%%%%%%%%%%
\footnote{The nontrivial potential $V(\phi)$ can allow for fragmentation of the condensate into particles~\cite{GarciaKe_2025}.  The couplings considered in this work are weak enough that this fragmentation is negligible during inflation (see \figref{quartic}).}
%%%%%%%%%%%%
\begin{equation}\label{eq:KG_condensate}
    \ddot\phi_0+3H_I\dot\phi_0+V'(\phi_0)=0,
\end{equation}
where dots represent derivatives with respect to $t$.  Let us suppose that the initial value $\phi_{0,i}$ satisfies $\alpha_i\gg1$.  In this case, the final term in \eqref{KG_condensate} dominates over the second term, so that $\phi_0$ oscillates in its potential.  The second term, however, damps its motion, leading to a redshifting of the energy density
\begin{equation}
    \rho\propto a^{-3(1+w)}=e^{-3(1+w)N},
\end{equation}
where $w$ is the equation of state of the spectator, which depends on its potential.  For instance, in a monomial potential
\begin{equation}\label{eq:potential}
    V(\phi)=\Lambda^{4-p}\phi^p,
\end{equation}
the equation of state is given by [see \eqref{eos_app}]
\begin{equation}\label{eq:eos}
    w=\frac{p-2}{p+2}.
\end{equation}
Noting that $\rho\propto V(p)$, this leads to a scaling
\begin{equation}\label{eq:alpha_exp}
    \alpha\propto\phi^{p-2}\propto e^{-\frac{6(p-2)}{p+2}N}
\end{equation}
during the initial stages of inflation.  In other words, if $\phi_0$ begins in a region where $\alpha$ is large, it will quickly evolve so that $\alpha$ decreases exponentially (as a function of $N$).

Let us now understand when $\phi_0$ begins to evolve slowly.   When the first term in \eqref{KG_condensate} can be neglected, the condensate follows the slow-roll solution
\begin{equation}\label{eq:slow_roll}
    \dot\phi_0=-\frac{V'(\phi_0)}{3H_I}.
\end{equation}
The slow-roll solution holds when $|\ddot\phi_0|\ll H|\dot\phi_0|$, or equivalently when $|\alpha|\ll1$.%
%%%%%%%%%%%%
\footnote{\label{ftnt:slowroll}%
If one allows for $H(t)$ to be time-dependent, one arrives at a second slow-roll condition from differentiating the denominator of \eqref{slow_roll}.  This condition is precisely the first slow-roll condition for inflation, $\epsilon\equiv-\dot H/H^2\ll1$.  Note that this is a condition on the inflaton sector, not the spectator potential.}
%%%%%%%%%%%%
The slow-roll solution in \eqref{slow_roll} can be re-expressed in terms of $\alpha$, as
\begin{equation}\label{eq:kappa}
    \frac{d\alpha}{dN}=-\kappa\alpha^2,\qquad\kappa\equiv\frac{V'''(\phi_0)V'(\phi_0)}{V''(\phi_0)^2}.
\end{equation}
For a monomial potential, $\kappa=\frac{p-2}{p-1}$.  This differential equation is solved by
\begin{equation}\label{eq:alpha_sol}
    \alpha(N)=\frac1{\kappa(N-N_\mathrm{sr})+1},
\end{equation}
where $\alpha(N_\mathrm{sr})\equiv1$.

We can now outline the scenario of interest.  Any potential $V(\phi)$ will consist of fast-roll regimes with $|\alpha|\gg1$ and slow-roll regimes with $|\alpha|\ll1$.  We are interested in the case where $\phi_0$ begins inflation in the fast-roll regime and rolls toward the slow-roll regime.  In such a scenario, $\alpha$ will decrease exponentially in $N$, so that $\phi_0$ quickly reaches the slow-roll regime.  From then on, it will decrease as $1/N$.  If inflation begins at $N_i$, then we generically expect
\begin{equation}\label{eq:alpha_est}
    \alpha(N)\approx\frac1{\kappa(N-N_i)}.
\end{equation}
See \citeR{McDonald_2003} for a similar scenario in the case where $\phi$ is a curvaton.  In the left plot of \figref{Pdel}, we show sample evolutions of $\alpha$ for two monomial potentials, which demonstrate this behavior.%
%%%%%%%%%%%%
\footnote{In this work, we ignore quantum fluctuations of $\phi$, i.e. we assume that its classical trajectory $\Delta\phi_\mathrm{cl}\sim\dot\phi/H_I$ dominates over its quantum fluctuations $\Delta\phi_\mathrm{qu}\sim H_I$~\cite{Dimopoulos_2003_pngb,Mazumdar_2011}.  For a monomial potential as in \eqref{potential}, this holds for $\alpha\gtrsim(\Lambda/H_I)^{(4-p)/(p-1)}$, while for a quartic potential as in \eqref{quartic} [with $m=0$], it holds for $\alpha\gtrsim\lambda^{1/3}$.  Note that the orange curve in the left plot of Fig.~1 violates this bound near the end of inflation (but this does not affect the right plot).}
%%%%%%%%%%%%

In summary, the necessary conditions on the potential and initial values for this scenario to occur are:
\begin{equation*}
    \begin{array}{cl}
    V''(\phi_{0,i})>3H_I^2&\text{(begin in fast-roll regime)}\\[2mm]
    V''(\phi)<3H_I^2\text{ for some }\phi&\text{(slow-roll regime exists)}\\[2mm]
    V'''(\phi)V'(\phi)>0&\text{(roll toward slow-roll regime)}\\[2mm]
    V(\phi_{0,i})\ll3H_I^2M_\mathrm{pl}^2&\text{(spectator subdominant)}
    \end{array}
\end{equation*}
Note that the third condition implies that this mechanism does not work for certain potentials, e.g. a cosine potential $V(\phi)\sim\cos(\phi)$.

%%%%%%%%%%%%%%%%%%%%%%%%%%%%%%%%%%%%%%%%%%%%%%%%%%%%%%%%%%%%%%%%%%%%%%%%%%%%%%%%%%%%%%%%%%
%%%%%%%%%%%%%%%%%%%%%%%%%%%%%%%%%%%%%%%%%%%%%%%%%%%%%%%%%%%%%%%%%%%%%%%%%%%%%%%%%%%%%%%%%%
\subsection{Perturbations}
\label{sec:field_perturbations}
%%%%%%%%%%%%%%%%%%%%%%%%%%%%%%%%%%%%%%%%%%%%%%%%%%%%%%%%%%%%%%%%%%%%%%%%%%%%%%%%%%%%%%%%%%
%%%%%%%%%%%%%%%%%%%%%%%%%%%%%%%%%%%%%%%%%%%%%%%%%%%%%%%%%%%%%%%%%%%%%%%%%%%%%%%%%%%%%%%%%%

Next we consider the perturbations $\delta\phi$ and compute the isocurvature power spectrum imprinted on the field during inflation.  In the absence of any potential, the resulting power spectrum is flat, while a nonzero $V''(\phi_0)$ imprints a blue tilt on the spectrum.  See \citeR[s]{Lyth_2002,Chung_2005,riotto2017,baumann2018,GarciaKe_2025} for similar derivations.  The perturbations of $\phi$ evolve as%
%%%%%%%%%%%%
\footnote{In the fast-roll regime $\alpha\gg1$, oscillations of the $V''(\phi_0)$ term in \eqref{delphik} can induce parametric resonance.  If this occurs at scale factor $a_p$, it will excite modes with $k/a_p\sim\sqrt{V''}\sim\sqrt{\alpha_p}H_I$.  For a monomial potential, one can show from \eqref{alpha_exp} that this mode has $\alpha_*\sim\alpha_p^{(8-2p)/(p+2)}$.  For $p\leq4$, this gives $\alpha_*\gtrsim1$, which corresponds to unobservable modes which are still outside the horizon.  However, for $p>4$, the effects of early parametric resonance may be imprinted on observable modes.}
%%%%%%%%%%%%
\begin{equation}\label{eq:delphik}
    \delta\ddot\phi_k+3H_I\delta\dot\phi_k+\left(\frac{k^2}{a^2}+V''(\phi_0)\right)\delta\phi_k=0,
\end{equation}
where $\delta\phi_k(t)$ are the Fourier modes of $\delta\phi(x,t)$.  This can be rewritten in terms of $f_k\equiv a\delta\phi_k$ and conformal time $\tau$ as
\begin{equation}\label{eq:perturbations}
    \partial_\tau^2f_k+\left(k^2-\frac{\partial_\tau^2a}a+a^2V''(\phi_0)\right)f_k=0.
\end{equation}

Before the mode exits the horizon, $|k\tau|\gg1$, \eqref{perturbations} takes the form of a harmonic oscillator with frequency $k$.  As initial conditions for this oscillator, we take the Bunch-Davies vacuum
\begin{equation}
    f_k(\tau\rightarrow-\infty)=\frac1{\sqrt{2k}}e^{-ik\tau}.
\end{equation}
In the superhorizon limit, $|k\tau|\ll1$, \eqref{perturbations} becomes
\begin{equation}
    \partial_\tau^2f_k+\frac{3\alpha(\tau)-2}{\tau^2}f_k=0,
\end{equation}
where $\alpha(\tau)$ encodes the evolution of the condensate $\phi_0(\tau)$.  For constant (or slowly varying) $\alpha\ll1$, this is solved by
\begin{equation}\label{eq:superhorizon}
    f_k\sim\frac{(-k\tau)^{\alpha-1}}{\sqrt k}\implies\delta\phi_k\sim H_I^{1-\alpha}k^{\alpha-\frac32}e^{-\alpha N}.
\end{equation}
Note that $\delta\phi_k$ decreases exponentially (as a function of $N$) while outside the horizon.  This means that longer wavelength modes, which exit the horizon earlier, will exhibit suppressed power compared to shorter wavelength modes.  In other words, this implies a blue tilt of the power spectrum.  Specifically, the power spectrum of the spectator perturbations (at the end of inflation) is approximately
\begin{align}
    \left.P_{\delta\phi}(k)\right|_\rh&\equiv\frac{k^3}{2\pi^2}|\delta\phi_k|^2\\&\approx\left(\frac{H_I}{2\pi}\right)^2\cdot\exp\left(-2\int_{N_*}^{N_\rh}\alpha(N)dN\right),
    \label{eq:Pdelphi}
\end{align}
where $N_*$ denotes horizon crossing ($k=a_*H_I$) and $N_\rh=0$ denotes the end of inflation.  The first factor in \eqref{Pdelphi} represents the well-known result for a free massless ($\alpha=0$) spectator~\cite{Birrell:1982ix,Baumann_2022}, while the second factor represents the damping due to the nonzero potential.

Ultimately, we are interested in the statistics of the density perturbations%
%%%%%%%%%%%%
\footnote{Here we assume that the energy density is dominated by the potential energy and neglect the kinetic and gradient energy contributions (see \eqref{rho} for full expression for $\rho$).}
%%%%%%%%%%%%
\begin{equation}
    \delta_k\equiv\frac{\delta\rho_k}\rho\approx\frac{V'(\phi_0)}{V(\phi_0)}\delta\phi_k,
\end{equation}
where $\delta\rho_k$ are the Fourier modes of $\rho(x,t)$.  From \eqref{Pdelphi}, we can compute the power spectrum of the density perturbations
\begin{equation}\label{eq:Pdel}
    \left.P_\delta(k)\right|_\rh\approx\frac1{12\pi^2\alpha_\rh}\left[\frac{V'^2V''}{V^2}\right]_\rh\exp\left(-2\int_{N_*}^{N_\rh}\alpha(N)dN\right).
\end{equation}
Importantly, $P_\delta$ scales with $k$ as%
%%%%%%%%%%%%
\footnote{For generic $\alpha$, \eqref{perturbations} is solved by $f_k\sim\sqrt{-\tau}H_\nu(-k\tau)$, where $H_\nu$ is a Hankel function (of either kind) of order $\nu=\sqrt{9/4-3\alpha}$, resulting in a tilt $n_\mathrm{iso}-1=3-2\nu$~\cite{riotto2017,baumann2018}.  \eqref[s]{superhorizon} and (\ref{eq:tilt}) are the $\alpha\ll1$ limits of these well-known results.}
%%%%%%%%%%%%
\begin{equation}\label{eq:tilt}
    n_\mathrm{iso}-1\equiv\frac{d\log P_\delta}{d\log k}=2\alpha(N_*).
\end{equation}

We see that the tilt of the spectrum is set by the value of $\alpha$ when the relevant mode exits the horizon.  As the modes we measure in the CMB exited the horizon around $N_*\approx-60$, then the tilt of the spectrum on large scales will be approximately $2/\kappa(N_\mathrm{tot}-60)$, where $N_\mathrm{tot}$ is the total number of $e$-folds during inflation.  For instance, for $\kappa=\mathcal O(1)$ and $N_\mathrm{tot}-60=\mathcal O(10)$, we expect $n_\mathrm{iso}-1=\mathcal O(0.1)$.  While this is a modest tilt, we note that it can exceed the measured tilt of the adiabatic fluctuations in the CMB~\cite{planck_inflation}.  For smaller $\kappa$ or fewer $e$-folds, $n_\mathrm{iso}-1$ may be larger.  In the right plot of \figref{Pdel}, we show the primordial isocurvature spectrum [with the constant factor from late-time evolution; see \secref{density_perturbations}] for two monomial potentials.  These are shown as a function of current momentum scale (see \appref{relic})
\begin{equation}\label{eq:ktoday}
    k_\mathrm{today}=\frac{a_\mathrm{rh}}{a_\mathrm{today}}k\approx10^{-28}\sqrt{\frac{10^{12}\,\mathrm{GeV}}{H_I}}\cdot k.
\end{equation}
The potential with small $\kappa$ yields large isocurvature fluctuations at small scales.

%%%%%%%%%%%%%%%%%%%%%%%%%%%%%%%%%%%%%%%%%%%%%%%%%%%%%%%%%%%%%%%%%%%%%%%%%%%%%%%%%%%%%%%%%%
%%%%%%%%%%%%%%%%%%%%%%%%%%%%%%%%%%%%%%%%%%%%%%%%%%%%%%%%%%%%%%%%%%%%%%%%%%%%%%%%%%%%%%%%%%
\section{Radiation Domination}
\label{sec:radiation}
%%%%%%%%%%%%%%%%%%%%%%%%%%%%%%%%%%%%%%%%%%%%%%%%%%%%%%%%%%%%%%%%%%%%%%%%%%%%%%%%%%%%%%%%%%
%%%%%%%%%%%%%%%%%%%%%%%%%%%%%%%%%%%%%%%%%%%%%%%%%%%%%%%%%%%%%%%%%%%%%%%%%%%%%%%%%%%%%%%%%%

%%%%%%%%%%%%%%%%%%%%%%%%%%%%%%%%%%%%%%%%%%
%%%%%%%%%%%%%%%%%%%%%%%%%%%%%%%%%%%%%%%%%%
\begin{figure}[t]
\includegraphics[width=0.99\columnwidth]{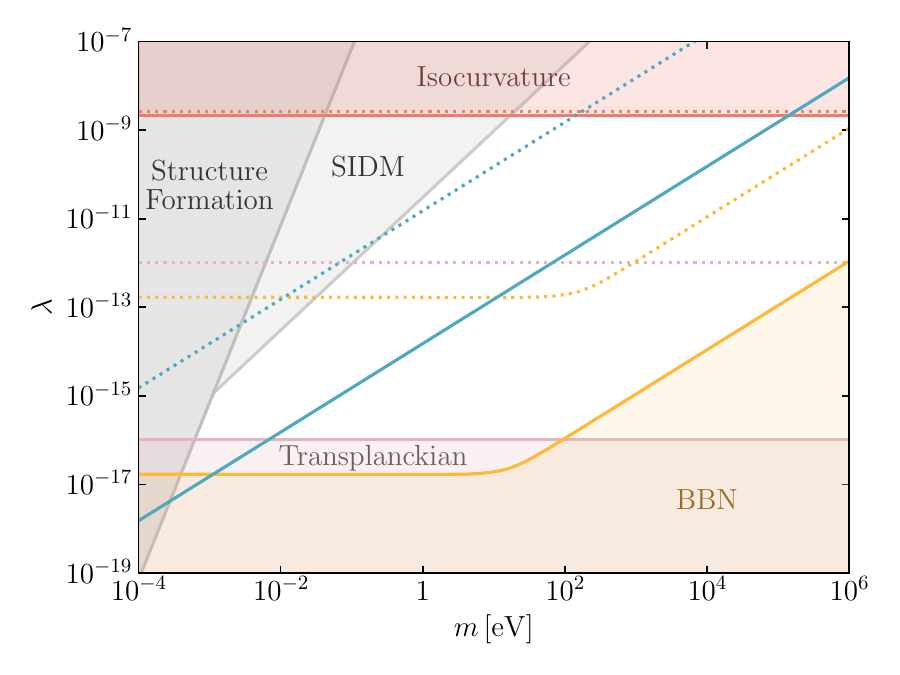}
\caption{\label{fig:quartic}%
    Parameter space for inflationary production of scalar DM with a quartic potential [see \eqref{quartic}].  The blue line indicates parameters which produce the correct DM relic abundance for $H_I=10^{10}\,\mathrm{GeV}$ and $N_\mathrm{tot}=80$ (assuming instantaneous reheating).  Colored regions indicate constraints on this production scenario, including: overproduction of isocurvature fluctuations (see right plot of \figref{Pdel}), violation of $N_\mathrm{eff}$ constraints at the time of BBN, and transplanckian initial conditions $\phi_{0,i}>M_\mathrm{pl}$.  (Note that all parameters below the blue line are also constrained due to overclosure of the universe.)  Dotted lines show these curves for $H_I=10^{12}\,\mathrm{GeV}$.  In grey, we show late-time constraints on self-interacting dark matter (SIDM) and disruption of structure formation~\cite{Budker_2023}.
    }
\end{figure}
%%%%%%%%%%%%%%%%%%%%%%%%%%%%%%%%%%%%%%%%%%
%%%%%%%%%%%%%%%%%%%%%%%%%%%%%%%%%%%%%%%%%%

Now that we have understood how the condensate and perturbations of the spectator field are generated during inflation, we study how these further evolve during radiation domination.  While in \secref{inflation}, we primarily dealt with the field $\phi$ itself, during radiation domination, it will be more useful to deal with the energy density $\rho$.  This is because $\phi_0$ begins to oscillate, and so if $V(\phi)$ is nontrivial, the final term in \eqref{delphik} can lead to parametric resonance of $\delta\phi_k$.  This implies that the perturbative decomposition of the field in \eqref{homogeneous} will break down.  The energy density will, however, remain nearly homogeneous on large scales.

For simplicity, in this section, we will assume instantaneous reheating at $t_\rh=0$ (or $\tau_\rh=-1/H_I$) into radiation domination.  This is described by the scale factor
\begin{equation}
    a=\sqrt{1+2H_It}=2+H_I\tau
\end{equation}
for $t>t_\rh$ (or $\tau>\tau_\rh$), which corresponds to a Hubble scale $H(a)=H_Ia^{-2}$.  In order to study the dynamics of the energy density, we will also need to fix a potential.  We will focus on a quartic potential with a small mass
\begin{equation}\label{eq:quartic}
    V(\phi)=\frac12m^2\phi^2+\frac\lambda{4!}\phi^4.
\end{equation}
We note that, for the parameter space of interest in this work, this potential is not technically natural, as the quartic term can lead to large radiative corrections for the mass term.  Nevertheless, similarly unnatural models are often considered for the inflaton or for DM~\cite{Arvanitaki_2015,planck_inflation,Batell_2023,Alachkar_2025}.

We will first study how the average energy density of the scalar evolves through radiation domination.  We will find viable parameter space where a scalar field with the potential in \eqref{quartic} can constitute the entirety of DM (see \figref{quartic}).  Then we will track the density perturbations and see that the primordial power spectrum is related to the power spectrum at reheating by an $\mathcal O(1)$ factor.  Therefore, the tilt derived in \eqref{tilt} remains valid at late times.

%%%%%%%%%%%%%%%%%%%%%%%%%%%%%%%%%%%%%%%%%%%%%%%%%%%%%%%%%%%%%%%%%%%%%%%%%%%%%%%%%%%%%%%%%%
%%%%%%%%%%%%%%%%%%%%%%%%%%%%%%%%%%%%%%%%%%%%%%%%%%%%%%%%%%%%%%%%%%%%%%%%%%%%%%%%%%%%%%%%%%
\subsection{Relic density}
\label{sec:relic}
%%%%%%%%%%%%%%%%%%%%%%%%%%%%%%%%%%%%%%%%%%%%%%%%%%%%%%%%%%%%%%%%%%%%%%%%%%%%%%%%%%%%%%%%%%
%%%%%%%%%%%%%%%%%%%%%%%%%%%%%%%%%%%%%%%%%%%%%%%%%%%%%%%%%%%%%%%%%%%%%%%%%%%%%%%%%%%%%%%%%%

First let us consider the evolution of the average energy density.  We saw in \secref{condensate} that, at the beginning of inflation, the scalar quickly reaches a slow-roll solution, which acts as an attractor.  Thus for a fixed length of inflation $N_\mathrm{tot}$, the energy density in the scalar at the end of inflation, and therefore also its energy density today, is independent of its initial conditions.  This is in contrast to other scalar DM production mechanisms where the field value at the end of inflation, and so also its abundance today, is stochastic~\cite{Garcia_2023,GarciaKe_2025}.  Enforcing the correct DM relic abundance then implies a predictive relationship between the parameters of the potential.

As radiation domination progresses, the scalar will exhibit various different equations of state, which will affect how its energy density redshifts.  The scalar will always begin radiation domination as it ended inflation: slow rolling.  This corresponds to an equation of state $w=-1$, or equivalently a constant energy density $\rho\propto a^0$.  If the scalar is to be the DM, it should end up redshifting like matter, i.e. $w=0$ or $\rho\propto a^{-3}$.  In the case of the potential in \eqref{quartic}, there will also be an intermediate period where the quartic term dominates.  In this case, \eqref{eos} implies that $w=1/3$ and so the scalar redshifts like radiation, $\rho\propto a^{-4}$, during this period.

More specifically, the scalar will end inflation with energy density
\begin{equation}\label{eq:rho_initial}
    \rho_\rh\approx V(\phi_\rh)=\frac{3\alpha_\rh^2H_I^4}{2\lambda}.
\end{equation}
It will begin to oscillate and behave as radiation at $a_r$ defined by
\begin{equation}
    \alpha_r\approx\frac{V''(\phi_\rh)}{3H(a_r)^2}\sim1\implies a_r\sim\alpha_\rh^{-1/4}.
\end{equation}
When the scalar behaves as radiation, the amplitude of the field redshifts as $\phi\sim a^{-1}$.  Finally, it will begin to behave as matter at $a_m$ defined by
\begin{equation}\label{eq:am}
    \frac12m^2\phi_m^2\sim\frac\lambda{4!}\phi_m^4\implies a_m\sim\sqrt[4]{\frac{\alpha_\rh}4}\frac{H_I}m.
\end{equation}
For $a>a_m$, the energy density in the scalar is therefore
\begin{equation}\label{eq:rho_final}
    \rho a^3\sim\rho_\rh\left(\frac{a_\rh}{a_r}\right)^0\left(\frac{a_r}{a_m}\right)^4a_m^3\sim\frac{3\alpha_\rh^{3/4}mH_I^3}{\sqrt2\lambda}.
\end{equation}
Setting this equal to the observed DM relic abundance fixes a relationship between $m$ and $\lambda$ (for fixed $H_I$ and $N_\mathrm{tot}$).

In \figref{quartic}, we show a precise calculation of this relationship (see \appref{relic} for details).  We also show other constraints on this production scenario.  If $\lambda$ is too large, the isocurvature fluctuations predicted in the right plot of \figref{Pdel} will contradict CMB observations.  On the other hand, if $\lambda$ is too small, the energy density in DM at the time of Big Bang Nucleosynthesis (BBN) will correspond to a large $\Delta N_\mathrm{eff}$, which is constrained to be $\Delta N_\mathrm{eff}<0.2$~\cite{Cyburt_2016}.  Additionally, if $\lambda$ is too small, the first condition at the end of \secref{condensate} would require $\phi_{0,i}>M_\mathrm{pl}$.  Finally, we also show late-time constraints on quartic DM.  Astrophysical observations constrain self interactions of DM to be $\sigma/m\lesssim1\,\mathrm{cm}^2/\mathrm{g}$~\cite{Adhikari_2022}.  Additionally, large quartic interactions can inhibit structure formation during matter domination~\cite{Budker_2023}.

%%%%%%%%%%%%%%%%%%%%%%%%%%%%%%%%%%%%%%%%%%%%%%%%%%%%%%%%%%%%%%%%%%%%%%%%%%%%%%%%%%%%%%%%%%
%%%%%%%%%%%%%%%%%%%%%%%%%%%%%%%%%%%%%%%%%%%%%%%%%%%%%%%%%%%%%%%%%%%%%%%%%%%%%%%%%%%%%%%%%%
\subsection{Density perturbations}
\label{sec:density_perturbations}
%%%%%%%%%%%%%%%%%%%%%%%%%%%%%%%%%%%%%%%%%%%%%%%%%%%%%%%%%%%%%%%%%%%%%%%%%%%%%%%%%%%%%%%%%%
%%%%%%%%%%%%%%%%%%%%%%%%%%%%%%%%%%%%%%%%%%%%%%%%%%%%%%%%%%%%%%%%%%%%%%%%%%%%%%%%%%%%%%%%%%

Finally, we study how the density perturbations that we derived in \eqref{Pdel} evolve through radiation domination.  To do so, we will apply a ``separate universes" argument~\cite{Wands_2000,Lyth_2005,Graham_2016,Artigas_2022}.  On large length scales, gradient terms in the equation of motion for the scalar can be ignored.  In this case, each point in space evolves independently of surrounding points.  In particular, each point will undergo the same dynamics that was outlined in \secref{relic} for the average energy density, only they will each begin with different local initial conditions.

The separate universes approach holds for momentum modes $k$ which are small enough that gradient terms have no effect on the dynamics of the density perturbations.  In \appref{perturbations}, we study the evolution of the density perturbations for a scalar field.  For a quartic potential, the effect of finite momentum $k$ is to induce oscillations of the perturbations at frequency $k/\sqrt3a$.  The separate universes approach therefore holds when the total accumulated phase
\begin{equation}
    \int dt\frac k{\sqrt3a}=\int da\frac{k}{\sqrt3H_I}
\end{equation}
is small.  Oscillations will cease at $a=a_m$, when the potential is no longer dominantly quartic.  We can therefore estimate that the separate universes approach holds for
\begin{align}\label{eq:sep_bound}
    k_\mathrm{today}&\lesssim\frac{a_\rh}{a_\mathrm{today}}\frac{\sqrt3H_I}{a_m}\sim1200\,\mathrm{Mpc}^{-1}\cdot\left(\frac{m}{10\,\mathrm{eV}}\right).
\end{align}

For scales satisfying this requirement, it is easy to see how the density perturbations evolve after reheating.  Each separate universe undergoes the dynamics outlined in \secref{relic} with a different initial energy density $\rho_\rh$, as well as a different local Hubble scale $H_I$.  The latter is due to local variations in the metric and can be related to the local gravitational potential $\Phi$ by $H_I\propto 1-\Phi$.  As such, density fluctuations caused by variations in $H_I$ are adiabatic perturbations, which correlate with the perturbations in all other components of the universe.  On the other hand, variations due to different initial $\rho_\rh$ are isocurvature perturbations, which only affect DM.  The dependence of the final energy density on these can easily be read off from \eqref[s]{rho_initial} and (\ref{eq:rho_final})%
%%%%%%%%%%%%
\footnote{It is not difficult to see that the final energy density will always have this dependence on $H_I$ [and therefore the same adiabatic perturbations in \eqref{primordial}], regardless of the potential.  This is because the ratio $a_r/a_m$ in \eqref{rho_final} depends only on $\rho_\rh$ and parameters of the potential.  In general, the amount of time spent progressing through different regions of the potential (with different equations of state) will not depend on $H_I$.  The Hubble scale $H_I$ will only affect the time at which the scalar begins to roll.}
%%%%%%%%%%%%
\begin{equation}
    \rho\propto\rho_\rh^{3/8}H_I^{3/2}.
\end{equation}
One then finds that the final density perturbations (for $a>a_m$) are
\begin{equation}\label{eq:primordial}
    \delta=\frac38\delta_\rh-\frac32\Phi.
\end{equation}
The former term represents isocurvature perturbations, while the latter represents adiabatic perturbations.  These are the total perturbations which constitute the primordial matter power spectrum [on scales which satisfy \eqref{sep_bound}].

%%%%%%%%%%%%%%%%%%%%%%%%%%%%%%%%%%%%%%%%%%%%%%%%%%%%%%%%%%%%%%%%%%%%%%%%%%%%%%%%%%%%%%%%%%
%%%%%%%%%%%%%%%%%%%%%%%%%%%%%%%%%%%%%%%%%%%%%%%%%%%%%%%%%%%%%%%%%%%%%%%%%%%%%%%%%%%%%%%%%%
\section{Conclusion}
\label{sec:conclusion}
%%%%%%%%%%%%%%%%%%%%%%%%%%%%%%%%%%%%%%%%%%%%%%%%%%%%%%%%%%%%%%%%%%%%%%%%%%%%%%%%%%%%%%%%%%
%%%%%%%%%%%%%%%%%%%%%%%%%%%%%%%%%%%%%%%%%%%%%%%%%%%%%%%%%%%%%%%%%%%%%%%%%%%%%%%%%%%%%%%%%%

In this work, we have shown that the dynamics of a scalar with a nontrivial potential during inflation can naturally give rise to a blue-tilted spectrum for its perturbations.  For a free scalar field, such a spectrum would typically require the mass to be comparable to the inflationary Hubble scale, $m\sim H_I$.  However, if the scalar has a potential $V(\phi)$ which satisfies the conditions at the end of \secref{condensate}, its condensate will spend most of inflation near the boundary of the slow-roll regime, defined by $\alpha=V''(\phi_0)/3H_I^2\sim1$.  As $V''(\phi_0)$ plays the role of an effective mass for the perturbations $\delta\phi$, then this is precisely the condition for a blue-tilted spectrum.  Such a spectrum can lead to interesting signatures at small scales, while avoiding CMB constraints at large scales.  A crucial condition for this mechanism is that inflation does not last too long, as the final blue tilt is inversely proportional to the number of $e$-folds before horizon crossing [see \eqref{alpha_est}].  This tilt can be enhanced for potentials with small $\kappa$, as in \eqref{kappa}.

We also highlighted that when $\phi$ is long-lived, this mechanism leads to a final relic abundance, which is insensitive to the initial conditions of the scalar field.  For fixed inflationary parameters, this implies a predictive relationship between the parameters of the potential in order to produce the correct relic abundance to be DM.  In \figref{quartic}, we computed the relationship between $m$ and $\lambda$ for a quartic potential, and plotted it in relation to constraints on this mechanism and quartic DM in general.  We demonstrated that there is viable parameter space where a scalar field with a quartic potential can constitute the entirety of DM.  If the DM contains no interactions with the SM, enhanced small-scale isocurvature (and associated signatures, such as gravitational waves or non-Gaussianities) may be one of the most promising ways of searching for it.

%%%%%%%%%%%%%%%%%%%%%%%%%%%%%%%%%%%%%%%%%%%%%%%%%%%%%%%%%%%%%%%%%%%%%%%%%%%%%%%%%%%%%%%%%%
%%%%%%%%%%%%%%%%%%%%%%%%%%%%%%%%%%%%%%%%%%%%%%%%%%%%%%%%%%%%%%%%%%%%%%%%%%%%%%%%%%%%%%%%%%
\acknowledgments
%%%%%%%%%%%%%%%%%%%%%%%%%%%%%%%%%%%%%%%%%%%%%%%%%%%%%%%%%%%%%%%%%%%%%%%%%%%%%%%%%%%%%%%%%%
%%%%%%%%%%%%%%%%%%%%%%%%%%%%%%%%%%%%%%%%%%%%%%%%%%%%%%%%%%%%%%%%%%%%%%%%%%%%%%%%%%%%%%%%%%

We thank David Kaplan, Wenqi Ke, Davide Racco, and Sarunas Verner for fruitful discussions on the mechanism presented in this work.  We thank Peizhi Du and Marco Gorghetto for discussions on existing constraints.  We also thank the Mainz Institute for Theoretical Physics and the organizers of the ``Windows into New Physics in the Sky" workshop, where valuable feedback on this work was received.

S.K. is supported in part by the DOE grant DE-SC0011842 and a Sloan Research Fellowship from the Alfred P. Sloan Foundation at the University of Minnesota.  IFAE is partially funded by the CERCA program of the Generalitat de Catalunya.  This work is supported by ERC grant ERC-2024-SYG 101167211 and is funded by the European Union.  Views and opinions expressed are however those of the author only and do not necessarily reflect those of the European Union or the European Research Council Executive Agency.  Neither the European Union nor the granting authority can be held responsible for them.

The code used for this research is made publicly available through Github~\cite{github} under CC-BY-NC-SA.

%%%%%%%%%%%%%%%%%%%%%%%%%%%%%%%%%%%%%%%%%%%%%%%%%%%%%%%%%%%%%%%%%%%%%%%%%%%%%%%%%%%%%%%%%%
%%%%%%%%%%%%%%%%%%%%%%%%%%%%%%%%%%%%%%%%%%%%%%%%%%%%%%%%%%%%%%%%%%%%%%%%%%%%%%%%%%%%%%%%%%
\bibliographystyle{JHEP}
\bibliography{references.bib}
%%%%%%%%%%%%%%%%%%%%%%%%%%%%%%%%%%%%%%%%%%%%%%%%%%%%%%%%%%%%%%%%%%%%%%%%%%%%%%%%%%%%%%%%%%
%%%%%%%%%%%%%%%%%%%%%%%%%%%%%%%%%%%%%%%%%%%%%%%%%%%%%%%%%%%%%%%%%%%%%%%%%%%%%%%%%%%%%%%%%%

\clearpage
\appendix

%%%%%%%%%%%%%%%%%%%%%%%%%%%%%%%%%%%%%%%%%%%%%%%%%%%%%%%%%%%%%%%%%%%%%%%%%%%%%%%%%%%%%%%%%%
%%%%%%%%%%%%%%%%%%%%%%%%%%%%%%%%%%%%%%%%%%%%%%%%%%%%%%%%%%%%%%%%%%%%%%%%%%%%%%%%%%%%%%%%%%
\renewcommand{\theequation}{A-\arabic{equation}}
\renewcommand{\thefigure}{A-\arabic{figure}}
\setcounter{equation}{0}
\setcounter{figure}{0}
\section{Relic density calculation}
\label{app:relic}
%%%%%%%%%%%%%%%%%%%%%%%%%%%%%%%%%%%%%%%%%%%%%%%%%%%%%%%%%%%%%%%%%%%%%%%%%%%%%%%%%%%%%%%%%%
%%%%%%%%%%%%%%%%%%%%%%%%%%%%%%%%%%%%%%%%%%%%%%%%%%%%%%%%%%%%%%%%%%%%%%%%%%%%%%%%%%%%%%%%%%

In this appendix, we outline the details for the computation of the relationship in \figref{quartic}.  A precise calculation of the relic density for the small masses appearing in \figref{quartic} is too computationally difficult, as it requires evolving over a very long period of time.  Instead, we compute the relic density for a much larger mass, and then extrapolate using the scalings derived in \eqref{rho_final}.  This relic density must then be connected to the current DM density.

To begin, the condensate can be evolved during inflation as a function of $t$ (or $N$), as in \eqref{KG_condensate}.  During radiation domination, the equation of motion for the condensate becomes
\begin{equation}
    \phi''_0+2a(\tau)H(\tau)\phi'_0+a(\tau)^2V'(\phi_0)=0,
\end{equation}
as a function of conformal time $\tau$.  Once the condensate begins rolling, the second term will become small, and the condensate will oscillate with frequency given by the final term.  Notice that when the quartic term in \eqref{quartic} dominates, this frequency is $\propto a^2\phi_0^2\sim a^0$.  In other words, when written in terms of $\tau$, the oscillation frequency is constant during the quartic phase.  For this reason, it is preferable to evolve as a function of $\tau$ (as opposed to $t$ or $N$) during radiation domination.

In order to get a robust prediction for the relic density, the condensate should be evolved until the quadratic term in \eqref{quartic} dominates.  (After this point, it will redshift as matter.)  For the masses in \figref{quartic}, this would require evolving to very large $a$, and therefore also large $\tau$.  Instead, we evolve the spectator for a much larger mass $m_\mathrm{ref}$, and compute the quantity $\rho a^3$ (which is constant at late times).  From \eqref{rho_final}, we can see this quantity scales as $\propto m^1H_I^3\lambda^{-1}$, and so can readily be extrapolated to other values of interest.

To connect this quantity with the current DM abundance, we must compute $a_\mathrm{today}$.  (Recall that our convention is $a_\mathrm{rh}=1$.)  This can be done by tracking the Hubble scale backwards in time to determine when it reaches $H_I$.  In the $\Lambda$CDM model, at late times, the Hubble constant is given by
\begin{equation}\label{eq:Hubble}
    H(a)=H_0\sqrt{\Omega_\Lambda+\Omega_m\cdot\frac{a_\mathrm{today}^3}{a^3}+\Omega_m\cdot\frac{a_\mathrm{eq}a_\mathrm{today}^3}{a^4}},
\end{equation}
where the final term represents the radiation component, normalized so that the matter and radiation components are equal at $a_\mathrm{eq}=a_\mathrm{today}/3400$~\cite{planck}.

At early times, the first two terms in \eqref{Hubble} can be neglected.  However, the radiation component evolves nontrivially when the number of SM degrees of freedom is changing.  The entropy density evolves as $s\propto g_{*,s}T^3\propto a^{-3}$, where $g_{*,s}$ is the entropy-weighted number of SM degrees of freedom.  The energy density therefore evolves as
\begin{equation}
    \rho\propto g_{*,\rho}T^4\propto g_{*,\rho}g_{*,s}^{-4/3}a^{-4},
\end{equation}
where $g_{*,\rho}$ is the energy-weighted number of degrees of freedom.  One can then see that, at early times, the Hubble scale is given by
\begin{equation}
    H(a)=\frac{g_{*,\rho}(a)^{1/2}g_{*,s}(a_\mathrm{today})^{2/3}}{g_{*,\rho}(a_\mathrm{today})^{1/2}g_{*,s}(a)^{2/3}}H_0\sqrt{\frac{a_\mathrm{eq}a_\mathrm{today}^3\Omega_m}{a^4}}.
\end{equation}
Setting $a=a_\mathrm{rh}=1$ and $H(a_\mathrm{rh})=H_I$, one can solve for $a_\mathrm{today}$.  Using $\Omega_m=0.32$ and $H_0=70\,\mathrm{km/s/Mpc}$~\cite{planck}, as well as the SM values for the degrees of freedom $g_{*,\rho}(a_\mathrm{today})=3.4$, $g_{*,s}(a_\mathrm{today})=3.9$, and $g_{*,s}(a_\mathrm{rh})=g_{*,\rho}(a_\mathrm{rh})=106.75$~\cite{Husdal_2016}, we find the result in \eqref{ktoday}.

The current DM abundance is given by
\begin{equation}
    \rho_\mathrm{DM}=\frac{3\Omega_cH_0^2}{8\pi G},
\end{equation}
where $\Omega_c=0.26$~\cite{planck} and $G$ is the gravitational constant.  The relationship between $m$ and $\lambda$ can then be determined by equating
\begin{equation}
    \rho_\mathrm{DM}a_\mathrm{today}^3=\left[\rho a^3\right]_\mathrm{ref}\left(\frac m{m_\mathrm{ref}}\right)\left(\frac{H_I}{H_{I,\mathrm{ref}}}\right)^3\left(\frac{\lambda_\mathrm{ref}}\lambda\right),
\end{equation}
where we compute $\left[\rho a^3\right]_\mathrm{ref}$ with the reference values $m_\mathrm{rm}=10^8\,\mathrm{GeV}$, $\lambda_\mathrm{ref}=10^{-9}$, and $H_{I,\mathrm{ref}}=10^{12}\,\mathrm{GeV}$.  (Note that $a_\mathrm{today}\propto H_I^{1/2}$, so ultimately $\lambda/m\propto H_I^{3/2}$.)

Finally, let us elaborate on the calculation of the BBN bound in \figref{quartic}.  This constraint is set by demanding that the DM abundance at the time of BBN has less energy density than $\Delta N_\mathrm{eff,max}=0.2$ neutrino species.  In order to determine the BBN abundance for parameter values of interest, we can again compute the abundance $\rho_\mathrm{ref}(a)$ over time for reference values and extrapolate.  From \eqref{rho_initial}, we see that the initial abundance scales as $\propto H_I^4\lambda^{-1}$.  The abundance then scales as $\rho\propto a^{-4}$, until $a_m\propto H_Im^{-1}$ [from \eqref{am}].  We can therefore relate the physical abundance to the reference calculation as
\begin{equation}
    \rho(a)=\rho_\mathrm{ref}\left(a\cdot\frac m{m_\mathrm{ref}}\cdot\frac{H_{I,\mathrm{ref}}}{H_I}\right)\left(\frac m{m_\mathrm{ref}}\right)^4\left(\frac{\lambda_\mathrm{ref}}\lambda\right).
\end{equation}
This must be less than
\begin{equation}
    \rho_\mathrm{BBN,max}=\frac{\pi^2}{30}\cdot\frac74\cdot\left(\frac4{11}\right)^{4/3}\cdot \Delta N_{\mathrm{eff,max}}\cdot T_\mathrm{BBN}^4
\end{equation}
at
\begin{equation}
    a_\mathrm{BBN}=a_\mathrm{today}\cdot\left(\frac{T_\mathrm{today}}{T_\mathrm{BBN}}\right)\cdot\left(\frac{g_{*,s}(a_\mathrm{today})}{g_{*,s}(a_\mathrm{BBN})}\right)^{1/3},
\end{equation}
where $T_\mathrm{BBN}=1\,\mathrm{MeV}$ and $g_{*,s}(a_\mathrm{BBN})=10.6$~\cite{Husdal_2016}.

%%%%%%%%%%%%%%%%%%%%%%%%%%%%%%%%%%%%%%%%%%%%%%%%%%%%%%%%%%%%%%%%%%%%%%%%%%%%%%%%%%%%%%%%%%
%%%%%%%%%%%%%%%%%%%%%%%%%%%%%%%%%%%%%%%%%%%%%%%%%%%%%%%%%%%%%%%%%%%%%%%%%%%%%%%%%%%%%%%%%%
\renewcommand{\theequation}{B-\arabic{equation}}
\renewcommand{\thefigure}{B-\arabic{figure}}
\setcounter{equation}{0}
\setcounter{figure}{0}
\section{Evolution of density perturbations}
\label{app:perturbations}
%%%%%%%%%%%%%%%%%%%%%%%%%%%%%%%%%%%%%%%%%%%%%%%%%%%%%%%%%%%%%%%%%%%%%%%%%%%%%%%%%%%%%%%%%%
%%%%%%%%%%%%%%%%%%%%%%%%%%%%%%%%%%%%%%%%%%%%%%%%%%%%%%%%%%%%%%%%%%%%%%%%%%%%%%%%%%%%%%%%%%

In this appendix, we derive the evolution of the density perturbations of a scalar field with nontrivial potential.  The dynamics of the field itself are governed by the Klein-Gordon equation in an expanding spacetime
\begin{equation}\label{eq:KG_full}
    \ddot\phi+3H\dot\phi-a^{-2}\nabla^2\phi+V(\phi)=0.
\end{equation}
If we multiply \eqref{KG_full} by $\dot\phi$, the result can be written as
\begin{equation}\label{eq:continuity}
    \dot\rho+3H(\rho+P)+\nabla\cdot\vec q=0,
\end{equation}
where
\begin{align}\label{eq:rho}
    \rho&=\frac12\dot\phi^2+\frac12a^{-2}\left(\nabla\phi\right)^2+V(\phi),\\
    P&=\frac12\dot\phi^2-\frac16a^{-2}\left(\nabla\phi\right)^2-V(\phi),\label{eq:pressure}\\
    \vec q&=-a^{-2}\dot\phi\nabla\phi.
\end{align}
\eqref{continuity} is the continuity equation for a fluid with energy density $\rho$, pressure $P$, and momentum density $\vec q$ in an expanding spacetime.  Similarly, if we multiply \eqref{KG_full} by $-a^{-2}\nabla\phi$, the result can be written as
\begin{equation}\label{eq:euler}
    \dot{\vec q}+5H\vec q+a^{-2}\nabla P+a^{-2}\nabla_j\Pi_{ij}=0,
\end{equation}
where
\begin{equation}
    \Pi_{ij}=a^{-2}\left(\nabla_i\phi\nabla_j\phi-\frac13(\nabla\phi)^2\delta_{ij}\right)
\end{equation}
is the anisotropic stress of the scalar field.  When written in terms of the velocity field, $\vec v=\vec q/(\rho+P)$, \eqref{euler} can be understood as the Euler equation for a fluid.

Let us make some simplifying assumptions for \eqref[s]{continuity} and (\ref{eq:euler}).  First, if the production of $\phi$ is isotropic, we can expect that $\Pi_{ij}\approx0$ on average.  Second, we would like to relate the energy density and pressure via an equation of state $P=w\rho$.  This can occur if all relevant modes are non-relativistic,%
%%%%%%%%%%%%
\footnote{More specifically, we require $k^2/a^2\ll\left\langle V''(\phi)\right\rangle$.  This is satisfied at the end of inflation for all modes well outside the horizon.  For $V(\phi)\sim\phi^p$ with $2\leq p\leq4$, the left-hand side decreases at least as fast as the right-hand side through radiation domination, and so this remains satisfied.}
%%%%%%%%%%%%
so that the second terms in \eqref[s]{rho} and (\ref{eq:pressure}) are negligible, and moreover if we are interested in timescales longer than the oscillation frequency.  In this case, the virial theorem implies
\begin{equation}
    \langle\dot\phi^2\rangle=\langle V'(\phi)\phi\rangle,
\end{equation}
where brackets represent an average over oscillations.  Note that this holds at any point in space.  Therefore, we can make the identification $\rho=wP$ locally with
\begin{equation}\label{eq:eos_app}
    ~~~~~~~~~~~~~~~~~~~~~~~~~~~~~w=\frac{\langle V'(\phi)\phi\rangle-2\langle V(\phi)\rangle}{\langle V'(\phi)\phi\rangle+2\langle V(\phi)\rangle}.~~~~~~~~~~~~~~~~
\end{equation}\vspace{2mm}

With these simpflications, \eqref[s]{continuity} and (\ref{eq:euler}) can be combined to form
\begin{equation}\label{eq:rho_ode}
    \ddot\rho+(8+3w)H\dot\rho+3(1+w)\left(\dot H+5H^2\right)\rho-wa^{-2}\nabla^2\rho=0.
\end{equation}
If we write \eqref{rho_ode} in terms of the density perturbations $\delta=\rho/\rho_0$, where $\rho_0\propto a^{-3(1+w)}$ is the average energy density, then we find
\begin{equation}\label{eq:del_ode}
    \ddot\delta+(2-3w)H\dot\delta-wa^{-2}\nabla^2\delta=0.
\end{equation}
When a mode of interest is outside the horizon, $k\ll aH$, then the final term in \eqref{del_ode} can be neglected.  Regardless of the equation of state $w$, the dominant solution for $\delta$ will then be constant.  In this case, the nontrivial potential has no effect on the evolution of the density perturbations.  However, once the mode has entered the horizon, $k\gg aH$, then it begins to oscillate at frequency $\sqrt w\cdot k/a$.  The dominant effect of a nontrivial potential is therefore to induce these oscillations once the mode enters the horizon.  In the case of scalar DM, these oscillations will continue until the scalar reaches the part of its potential which is dominantly quadratic $V(\phi)\sim\phi^2$ (at which point $w=0$).

%%%%%%%%%%%%%%%%%%%%%%%%%%%%%%%%%%%%%%%%%%%%%%%%%%%%%%%%%%%%%%%%%%%%%%%%%%%%%%%%%%%%%%%%%%
%%%%%%%%%%%%%%%%%%%%%%%%%%%%%%%%%%%%%%%%%%%%%%%%%%%%%%%%%%%%%%%%%%%%%%%%%%%%%%%%%%%%%%%%%%
%%%%%%%%%%%%%%%%%%%%%%%%%%%%%%%%%%%%%%%%%%%%%%%%%%%%%%%%%%%%%%%%%%%%%%%%%%%%%%%%%%%%%%%%%%
%%%%%%%%%%%%%%%%%%%%%%%%%%%%%%%%%%%%%%%%%%%%%%%%%%%%%%%%%%%%%%%%%%%%%%%%%%%%%%%%%%%%%%%%%%
\end{document}